\documentstyle[aps,psfig]{revtex}

\begin{document}

\twocolumn[\hsize\textwidth\columnwidth\hsize\csname 
@twocolumnfalse\endcsname 

\title{Dual equivalence between Self-Dual and Maxwell-Chern-Simons models\\
coupled to dynamical U(1) charged matter}

\author{M. A. Anacleto$^1$, A. Ilha$^2$, J. R. S. Nascimento$^1$,
R. F.Ribeiro$^1$ and C. Wotzasek$^2$}

\address{$^1$Departamento de F\' \i sica, Universidade Federal da
Para\'\i ba\\ 
58051-970 Jo\~ao Pessoa, Para\'\i ba, Brasil\\
$^2$Instituto de F\'\i sica, Universidade
Federal do Rio de Janeiro\\
21945-970, Rio de Janeiro, Brazil}


\maketitle

\begin{abstract}
We study the equivalence between the self-dual and the Maxwell-Chern-Simons
(MCS) models coupled to dynamical, U(1) charged matter, both fermionic and
bosonic. This is done through an iterative procedure of gauge
embedding that produces the dual mapping of the self-dual vector field
theory into a Maxwell-Chern-Simons version.
In both cases, to establish this equivalence a current-current interaction
term is needed to render the matter sector unchanged.  Moreover, the minimal
coupling of the  original self-dual model is replaced by a non-minimal
magnetic like coupling in the MCS side. Unlike the fermionic instance
however, in the bosonic example the dual mapping proposed here leads
to a Maxwell-Chern-Simons theory immersed in a field dependent medium.
\end{abstract}
\vskip2pc]

\section{Introduction}

The essential features manifested by the three dimensional field theories,
such as parity breaking and anomalous spin, are basically connected to the
presence of the topological and gauge invariant Chern-Simons
term (CST) \cite{CST}.
Three dimensional models with CST have provided deep insights in unrelated
areas in particle physics and condensed matter, both from the theoretical
and phenomenological points of view \cite{DJT}.
In this regard we mention the high temperature asymptotic of four dimensional
field theory models and the understanding of the universal behavior of the
Hall conductance in interacting electron systems. In particular this result
has been of great significance in order to extend the bosonization program
from two to three dimensions with important phenomenological
consequences \cite{boson}.

What is worthy of discussion, in this context, is that in three dimensions
there are two different ways to describe the dynamics of a single, freely
propagating spin one massive mode.  To that purpose, as first shown by Deser
and Jackiw \cite{DJ}, one can use either the self-dual (SD)
theory \cite{TPvN},

\begin{equation}
\label{PB10}
{\cal L}_{SD} = \frac 12 m^2 f_\mu f^\mu - \frac m2 
\varepsilon ^{\mu\nu\rho} f_\mu\partial_\nu f_\rho \; ,
\end{equation}
or the Maxwell-Chern-Simons (MCS) theory,

\begin{equation}
\label{PB20}
{\cal L}_{MCS} = -\frac 14 F_{\mu\nu} F^{\mu\nu} + \frac 14 m
 \varepsilon ^{\mu\nu\rho} A_\mu F_{\nu\rho}\; , 
\end{equation}
where $ F_{\mu\nu}= \partial_\mu A_\nu - \partial_\nu A_\mu$.
Using the master action concept \cite{DJ}, it was shown that in fact these
models are duality related to each other and established the identification

\begin{equation}
\label{PB30}
f^\mu \leftrightarrow F^\mu = \frac{1}{ m} \varepsilon ^{\mu\nu\rho}
\partial_\nu A_\rho
\end{equation}
that relates the basic field of the SD model with the dual of the MCS field.
This correspondence displays the way the gauge symmetry of the MCS representation,
$A_\mu\to A\mu + \partial_\mu \epsilon$, gets hidden in the SD representation, leading to
$f_\mu \to f_\mu$.

This well established equivalence between the SD and the MCS theories is
maintained even when the vector fields are coupled to external sources.
In fact such a coupling,

\begin{eqnarray}
\label{PB33}
{\cal L}_{SD} & \to & {\cal L}_{SD} + f_\mu J^\mu\nonumber\\
{\cal L}_{MCS} & \to & {\cal L}_{MCS} + A_\mu G^\mu
\end{eqnarray}
will not change this picture as long as one extends the identifications to
the sources as well.
A direct  inspection of the equations of
motion for the self-dual field $f^\mu$,
\begin{eqnarray}
\label{PB35}
-m\varepsilon^{\nu\alpha\beta}\partial_\alpha f_\beta + m^2 f^\nu &=& e
J^\nu 
\end{eqnarray}
and the MCS field $A^\mu$,

\begin{eqnarray}
\label{PB36}
-m\varepsilon^{\nu\alpha\beta}\partial_\alpha F_\beta + m^2 F^\nu &=& e
G^\nu\; .
\end{eqnarray}
shows that the models are classically equivalents if the extended
identification
\begin{eqnarray}
\label{6}
f^\mu \leftrightarrow F^\mu \Longrightarrow J^\mu \leftrightarrow
G^\mu \; ,
\end{eqnarray}
is made. Moreover, it has been shown recently by Gomes et al. \cite{GMdS} that such equivalence can also be successfully extended to the case of couplings with dynamical fermionic matter fields under some special circumstances. 

However, in regards to the dualization process in general, it should be observed that not all the problems have been solved successfully. For instance, in the case of non-abelian symmetries, such extension has been tried in the context of the master action and it has been observed in \cite{BFMS} that, apart from the case of weak coupling constant, this equivalence, alas, fails. What is more important for us at this juncture, it has been mentioned in \cite{GMdS} that the coupling with bosonic dynamical sources seems to run into trouble since in this case the gauge currents will have explicit dependence on the gauge field implying that the corresponding functional determinant will have a non-trivial dependence on the matter fields, unlike the fermionic case.

In a recent report \cite{IW} an alternative route to establish dual
equivalences between gauge and non-gauge theories has been proposed
that is based on the local lifting of the global symmetries present in
the non-gauge action.
This is done by iteratively incorporating counter-terms
into the action depending on powers of the
Euler vectors\cite{footnote} along with a set of ancillary fields.
Clearly the resulting embedded theory is dynamically equivalent to the original one.
This gauge embedding approach has been applied to
the case of the three dimensional non-abelian self-dual action which was
proved to be equivalent to the Yang-Mills-Chern-Simons (YMCS) theory for
the full range of values of the coupling constant and not only on the weak
coupling regime.

This alternative approach to dual transformation that is dimensionally
independent and sufficiently general to encompass both abelian and
non-abelian symmetries is here used to study the dual equivalence of
the SD and MCS theories for the case of dynamical couplings with both
fermionic and bosonic matter fields.  This is done next in section II.
The fermionic case discussed by Gomes, Malacarne and da Silva is
considered first in subsection II.A where the results of \cite{GMdS}
are reobtained and the dual mapping (\ref{6}) is analyzed
under this new point of view.  It is clearly shown that the dual operation,
as defined here, transforms the minimal coupling of the SD model into a
non-minimal magnetic coupling for the MCS case,

\begin{equation}
\label{PB07}
f_\mu J^\mu \rightarrow  A_\mu G^\mu,
\end{equation}
where 

\begin{equation}
\label{PB08}
 G^\mu= \frac 1m \varepsilon ^{\alpha\beta\mu} \partial_\alpha
J_\beta,
\end{equation}
and automatically provides for the presence of the Thirring like current-current interaction. In Ref.\cite{GMdS} this term is included, in an iterative trial process, in order to render the fermionic sectors in both sides identical.
In subsection II.B, that contains our main result, we study the case of minimal coupling of the SD theory with a dynamical, U(1) charged bosonic theory.  Section III is reserved to discuss this equivalence and to present our conclusions.

\section{The gauge embedding procedure}

The existence of gauge invariance in systems with second class constraints
has experienced a spate of interest in recent times \cite{vithee1}.
Basically this involves disclosing, in the language of constraints, hidden
gauge symmetries in such systems.
This situation may be of usefulness since one can consider the non-invariant
model as the gauge fixed version of a gauge theory.
The former reverts to the latter under certain gauge fixing conditions.
By doing so it has sometimes been possible to obtain a deeper and more
illuminating interpretation of these systems.
The associate gauge theory is therefore to be considered as the ``gauge
embedded" version of the original second-class theory. The advantage in
having a gauge theory lies in the fact that the underlying gauge symmetry
allows us to establish a chain of equivalence among different models by
choosing different gauge fixing conditions.

In this section we apply an iterative procedure to construct a
gauge invariant theory out of the self-dual model coupled to dynamical
matter fields, either fermionic or bosonic.
To guarantee equivalence with the starting non invariant theory we only
use counter-terms vanishing in the space of the solutions of the model.

\subsection{The fermionic case}

To present the basic principles and also to establish our notation, let us
consider first a Dirac field minimally coupled to a vector field specified
by the SD model \cite{GMdS}, so that the Lagrangian becomes

\begin{equation}
\label{PB40}
{\cal L}_{min}^{(0)} = {\cal L}_{SD}  - e f_\mu J^\mu + 
{\cal L}_{D}  \; , 
\end{equation}
where $J^\mu = \bar{\psi}\gamma^\mu \psi$, $M$ is the fermion mass and the
superscript index is an iterative counter. Here the Dirac Lagrangian is,

\begin{equation}
\label{PB45}
{\cal L}_{D} =  \bar{\psi}(i\partial\!\!\! /  -M)\psi \; , 
\end{equation}

Our basic goal is to transform the hidden symmetry of the Lagrangian (\ref{PB40})
into a local gauge symmetry

\begin{equation}
\label{PB47}
\delta f_\mu = \partial_\mu \epsilon
\end{equation}
with the lift of the global parameter into its local form, i.e.

\begin{equation}
\label{PB48}
\epsilon \to \epsilon(x,y,t)
\end{equation}

A variation of the Lagrangian (\ref{PB40}) gives the Euler vector as,

\begin{equation}
\label{PB50}
K^\mu = m^2 f^\mu - m \epsilon^{\mu\nu\lambda}\partial_\nu
f_\lambda - e J^\mu
\end{equation}
whose kernel gives the equations of motion of the model.

We mention at this juncture that an (weakly) equivalent description of this theory
may be obtained by adding to the original Lagrangian (\ref{PB40}) any function of the Euler vectors

\begin{equation}
\label{PB51}
{\cal L}_{min}^{(0)} \to {\cal L}_{min}^{(0)} + f(K_\mu)
\end{equation}
such that it vanishes at the space of solutions of (\ref{PB40}), i.e., $f(0)=0$.
To find the specific form of this function that also induces a gauge symmetry into ${\cal L}_{min}^{(0)}$
we work out iteratively.  To this end we define a first-iterated Lagrangian as,

\begin{equation}
\label{PB60}
{\cal L}_{min}^{(1)} = {\cal L}_{min}^{(0)} - B_\mu K^\mu\, .
\end{equation}
where the Euler vector has been imposed as a constraint with $B_\mu$ acting as a Lagrange multiplier.

The transformation properties of $B_\mu$ accompanying the basic field transformation (\ref{PB47})
is chosen so as to cancel the variation of ${\cal L}_{min}^{(0)}$, which gives

\begin{equation}
\label{PB70}
\delta B_\mu = \partial_\mu \epsilon
\end{equation}
A simple algebra then shows

\begin{eqnarray}
\label{PB80}
\delta {\cal L}_{min}^{(1)} &=&  - B_\mu \delta K^\mu\nonumber\\
&=& - \delta \left(\frac {m^2}2 B^2\right) + m B_\mu \epsilon^{\mu\nu\lambda}
\partial_\nu \delta f_\lambda
\end{eqnarray}
where we have used (\ref{PB47}) and (\ref{PB70}).  Because of (\ref{PB47}), the
second term in the r.h.s. of (\ref{PB80}) vanishes identically leading to a
second iterated Lagrangian,

\begin{equation}
\label{PB90}
{\cal L}_{min}^{(2)} = {\cal L}_{min}^{(1)} + \frac {m^2}2 B^2
\end{equation}
that is gauge invariant under the combined local transformation of $f_\mu$ and $B_\mu$.

We have therefore succeed in transforming the global SD theory into a locally
invariant gauge theory.  We may now take advantage of the Gaussian
character of the auxiliary field $B_\mu$ to rewrite (\ref{PB90})
as an effective action depending only on the original variable.
To this end we solve (\ref{PB90}) for the field $B_\mu$ (call this solution $\bar B_\mu$)
and replace it back into (\ref{PB90}) to find,

\begin{eqnarray}
\label{PB100}
& &{\cal L}_{eff} = {\cal L}_{min}^{(2)}\mid_{B_\mu = \bar B_\mu} \;=
{\cal L}_{min}^{(0)} - \frac 1{2m^2} K^2 \nonumber\\
& & = \frac 12 m^2 A_\mu A^\mu - \frac m2 
\varepsilon ^{\mu\nu\rho} A_\mu\partial_\nu A_\rho -e A_\mu J^\mu + {\cal L}_{D}\nonumber\\
& &-\frac 1{2m^2} \left[m^2 A^\mu - m \epsilon^{\mu\nu\lambda}
\partial_\nu A_\lambda - e J^\mu\right]^2
\end{eqnarray}
where we have used the 
structure of the Euler vector (\ref{PB50}) and renamed the basic fields $f_\mu \to A_\mu$ to reflect the 
embedded gauge invariance of the Lagrangian (\ref{PB100}). A further manipulation gives,

\begin{eqnarray}
\label{PB110}
{\cal L}_{eff} = {\cal L}_{MCS}  + 
\bar{\psi}(i\partial\!\!\! /  -M)\psi - \frac {e^2}{2 m^2} J_\mu J^\mu
- e A_\mu G^\mu
\end{eqnarray}
which is the result of \cite{GMdS}. It is noteworthy the presence of the
Thirring like interaction that appears naturally as a consequence of the
embedding algorithm and the non-minimal magnetic interaction in the MCS side.
This comes around since the magnetic contribution to be included in the
covariant derivative reads,

\begin{eqnarray}
\bar{\psi} \sigma^{\mu\nu}\psi\, F_{\mu\nu} =
\frac 12 \epsilon^{\mu\alpha\beta} J_\mu F_{\alpha\beta}
\sim A^\mu G_\mu
\end{eqnarray}
which results from the dual mapping of the minimal interaction term as
shown in (\ref{6}) and (\ref{PB08}).

The Thirring like term is fundamental to maintain the contents of the
fermionic sectors unchanged \cite{GMdS}. This can be seen by carefully
examining the dynamics of the fermionic sectors for both theories. 
The fermionic equation for the MCS sector reads,

\begin{eqnarray}
\label{AIS05234}
\left(i\,\slash\!\!\!{\partial} - M\right)\psi =
\left[e F^\mu + \frac{e^2}{m^2} J^\mu\right]\,\gamma_{\mu}\,\psi\,,
\end{eqnarray}
To eliminate the bosonic MCS field $F^\mu$, let us rewrite the equation
of motion for the MCS field (\ref{PB36}) as,

\begin{eqnarray}
\label{AIS111}
R_{\mu\nu}^{-1} F^{\nu} = e\, G_{\mu}\,,
\end{eqnarray}
where $R_{\mu\nu}$ is a differential operator such that its inverse is 
defined as
\begin{eqnarray}
\label{AIS08}
R^{-1}_{\mu\nu} = -m\,\epsilon_{\mu\lambda\nu}\,\partial^{\lambda} + 
m^{2}\,\eta_{\mu\nu}\,.
\end{eqnarray}
Substituting this equation back in the matter equation gives,

\begin{eqnarray}
\label{AIS05235}
\left(i\,\slash\!\!\!{\partial} - M\right)\psi = 
\left[e^2 R^{\mu\nu}G_\nu + \frac{e^2}{m^2}
J^\mu\right]\,\gamma_{\mu}\,\psi\,,
\end{eqnarray}
that is now written completely in terms of the fermionic fields.

{}From the self-dual model (\ref{PB40}) we find the equations of motion
for the fermion $\bar{\psi}$ as,

\begin{eqnarray}
\label{AIS0523}
\left(i\,\slash\!\!\!{\partial} - M\right)\psi = 
e\,f^{\mu}\,\gamma_{\mu}\,\psi\, .
\end{eqnarray}
Using the differential operator $R^{\mu\nu}$, the self-dual equation of
motion (\ref{PB35}) now reads,

\begin{eqnarray}
\label{AIS1115}
R_{\mu\nu}^{-1} f^{\nu} = e\, J_{\mu}\,,
\end{eqnarray}
Using next that $R^{\mu\nu}R_{\nu\lambda}^{-1}=\delta^\mu_\lambda$
and (\ref{AIS1115}) we can rewrite (\ref{AIS0523}) as,

\begin{eqnarray}
\label{AIS0567}
\left(i\,\slash\!\!\!{\partial} - M\right)\psi &=& e^2\,R^{\mu\nu}\,J_\nu\,
\gamma_{\mu}\,\psi \nonumber\\
&=& \left[e^2 R^{\mu\nu} G_\nu + \frac{e^2}{m^2} J^\mu\right]\,
\gamma_{\mu}\,\psi 
\end{eqnarray}
that equals (\ref{AIS05235}) showing that the fermionic sector of both
theories have the same dynamics.

We may now exam the dual mapping $f_\mu \to F_\mu$ for the case
of coupling with dynamical fermions. From the MCS equation of
motion (\ref{PB36}) and the definition of the magnetic charge we have,

\begin{eqnarray}
F^\mu &=& \frac em R^{\mu\nu} \epsilon_{\nu\lambda\rho}
\partial^\lambda J^\rho \nonumber\\
&=& e R^{\mu\nu} J_\nu - \frac 2{m^2} J^\mu
\end{eqnarray}
where we have used (\ref{AIS08}) and the identity
$R^{\mu\nu}R_{\nu\lambda}^{-1}=\delta^\mu_\lambda$.  From the equation
of motion for the self-dual field (\ref{PB35}) we finally obtain that
the correct map from self-dual to MCS models when dynamical charged
fermionic matter is present is given by,

\begin{eqnarray}
\label{mapping}
f^\mu \to F^\mu + \frac e{m^2} J^\mu\, ,
\end{eqnarray}
that  does not correspond to the initial guess 
$f_{\mu} \to F_{\mu}$ which is valid only for the free case. This proves
that alone, the minimal self-dual model
does not correspond to the magnetically coupled Maxwell Chern-Simons model.
The solution is found to be an additional Thirring like 
term. Equivalently, one may add a 
minimal coupling in the master action proposed by Deser and Jackiw,

\begin{eqnarray}
\label{AIS13}
{\cal L} = \frac{m^{2}}{2}\,f^{\mu}\,f_{\mu} - m^{2}\,f^{\mu}\,F_{\mu} 
+ \frac{m^{2}}{2}\,F^{\mu}\,A_{\mu} - e\,J^{\mu}\,f_{\mu}\,,
\end{eqnarray}
so as to compensate for the additional current term. One may check that
indeed this master action produces the correct mapping (\ref{mapping}).

\subsection{The bosonic case}

Let us consider next the case of bosonic matter that is seen to be difficult
to solve through the traditional approach of master Lagrangian \cite{GMdS}.
The self-dual model minimally coupled to U(1) charged bosonic matter is
described by the following Lagrangian density,

\begin{equation}
\label{PB120}
{\cal{L}}_{min}^{(0)}={\cal L}_{SD} + {\cal L}_{int}+{\cal{L}}_{KG},
\end{equation}
where

\begin{eqnarray}
{\cal L}_{SD} &=& \frac{m^{2}}{2}f^{\mu}f_{\mu}
-\frac{m}{2}\varepsilon^{\mu\nu\rho}f_{\mu}\partial_{\nu}f_{\rho}\nonumber\\
{\cal L}_{int} &=& -ef_{\mu}J^{\mu}+e^{2}f^{\mu}f_{\mu}\phi^{*}\phi\nonumber\\
{\cal{L}}_{KG}&=&\partial_{\mu}\phi^{*}\partial^{\mu}\phi-M^{2}\phi^{*}\phi,
\, ,
\end{eqnarray}
are the self-dual, interaction and Klein-Gordon Lagrangians for a vector
field and a massive U(1) charged, complex scalar field, respectively.
Here,

\begin{equation}
\label{PB130}
J^{\mu}=i(\phi^{\ast}\partial^{\mu}\phi-\partial^{\mu}\phi^{\ast}\phi),
\end{equation}
is the global Noether current associated to a U(1) phase transformation.
Notice that the gauge current, obtained from the interaction Lagrangian as,

\begin{equation}
\label{PB138}
{\cal J}^\mu = \frac 1e\frac{\delta {\cal{L}}_{int}}{\delta f_\mu} =
J_\mu - 2 e \left(\phi^{*}\phi\right) f_\mu
\end{equation}
is explicitly dependent on the vector field which, as mentioned in the
introduction, makes matters complicated.

It is interesting at this instance to compute the field equations for the
vector, self-dual field $f^\mu$ and for the scalar field $\phi$.
We find, for the vector field

\begin{eqnarray}
\label{LL05}
\left(\mu^2 \eta_{\mu\alpha} - m \epsilon_{\mu\rho\alpha}\partial^\rho\right)
f^\alpha = e\,  J_\mu\, ,
\end{eqnarray}
that can be solved as,

\begin{eqnarray}
\label{LL10}
f^\alpha = e\, m\, M^{\alpha \beta}  J_\beta\, .
\end{eqnarray}
The field operator $ M_{\alpha\beta}$ is defined in a way, such that its
inverse is given as,

\begin{eqnarray}
\label{LL30}
M_{\alpha \beta}^{-1} = m\,\mu^2 \eta_{\alpha\beta} -  {m^2} \,
\epsilon_{\alpha\beta\rho}\partial^\rho 
\end{eqnarray}
and $\mu$ is a ``field dependent'' mass parameter defined as,

\begin{equation}
\label{PB150}
\mu^2 = m^2 + 2 e^2  \phi^{*}\phi\, ,
\end{equation}
that is unlike (\ref{AIS08}) and (\ref{AIS1115}) defined for the fermionic
case which are not field dependent.
For the scalar field we find,

\begin{eqnarray}
\label{LL20}
& &\left(\partial^\mu\partial_\mu  + M^2\right) \phi = e^2 \,
f_\mu f^\mu \, \phi - 2 i\, e\, \partial^\mu\phi\, f_\mu\nonumber\\
& & = m^2\, e^4\, \phi\, \left( M_{\alpha\beta}  J^\beta\right)^2
- 2\, i\, m\,e^2\,\partial^\alpha \phi\, M_{\alpha\beta} J^\beta
\end{eqnarray}

Following the procedure already outlined in the fermionic case, we compute
the Euler vector,

\begin{equation}
\label{PB140}
K^\mu = \mu^2 f^\mu - m \epsilon^{\mu\nu\lambda}\partial_\nu f_\lambda -
e J^\mu 
\end{equation}
As usual, we introduce the ancillary field $B_\mu$ as a Lagrange multiplier
to impose this relation weakly on the theory. Next we embed a local gauge symmetry,
leading to the final gauge invariant action as,

\begin{equation}
\label{PB155}
{\cal L}_{min}^{(2)} = {\cal L}_{min}^{(0)} - B_\mu K^\mu + \frac {m^2}2 B^2
\end{equation}
Next, we get rid of the ancillary field $B_\mu$ using its equations of motion,

\begin{equation}
\label{PB160}
 B^\mu =\frac 1{\mu^2}\, K^\mu
\end{equation}
yielding the effective action,

\begin{eqnarray}
\label{PB170}
{\cal L}_{eff}&=&{\cal L}_{KG}- \frac{m^2}{4\,\mu^2}\, F_{\mu\nu}^2 +
\frac m2\,\epsilon_{\alpha\beta\rho}\, A^\alpha\partial^\beta
A^\rho\nonumber\\
& &-\frac {e^2}{2 \mu^2} J^2 -\frac{e\,m^2}{\mu^2}F^\alpha J_\alpha\, ,
\end{eqnarray}
where, as before, we have renamed $f_\mu \to A_\mu$ to stress the invariant
character of the theory. Notice that this result displays the same structure
as the fermionic case with the minimal coupling being replaced by a
non-minimal magnetic
coupling and the presence of the Thirring like current-current piece.
The important difference is that now the coefficient of the Maxwell term
is field dependent.
This structure is known to appear in the abelian Higgs model with an
anomalous magnetic interaction leading to an effective model with field
dependent permeability \cite{P}. This comes around if we use a generalized
covariant derivative given as,

\begin{eqnarray}
{\cal D}_\mu \phi = \partial_\mu \phi -
i\, e\, A_\mu \phi - \, \frac i4 \gamma \epsilon_{\mu\alpha\beta}\,
F^{\alpha\beta} \phi
\end{eqnarray}
where $\gamma$ is the strength of the magnetic moment. It naturally leads
to a term as $\frac {\gamma}2 J_\mu \epsilon^{\mu\alpha\beta}
\partial_\alpha A_\beta$ which is proportional to the last term
in (\ref{PB170}).
It is interesting to observe that in this instance the Thirring like
interaction is not enough to guarantee the invariance of the contents of
the matter sector,
unlike the fermionic case, but additional pieces coming from the field
dependent mass becomes necessary.  This is in contrast with the fermionic
case and it is basically tied to the second-order character of the bosonic
action and the field dependence of the gauge current (\ref{PB138}).

It is mandatory to check if the matter sector of the dual
version (\ref{PB170}) is in agreement with the corresponding matter
sector of the original self-dual model (\ref{PB120}).  To this end we
work out the field equations both for the vector and the scalar fields
derived from (\ref{PB170}).  For the first we find, after some algebra,

\begin{eqnarray}
\label{LL100}
\frac 1{\mu^2}\,F^\alpha &=& e\,
M^{\alpha \beta}\epsilon_{\beta\rho\sigma}\partial^\rho 
\left(\frac 1{\mu^2} J^\sigma\right)\nonumber\\
&=& \frac {e}{m} M^{\alpha \beta}J_\beta - \frac{e}{m^2\,\mu^2} J^\alpha
\end{eqnarray}
where we have used the property

\begin{eqnarray}
\label{LL110}
 M^{\alpha \beta}\epsilon_{\beta\mu\nu}\partial^\mu \left(\frac 1{\mu^2}
J^\nu\right)=\frac{1}{m} M^\alpha_\nu J^\nu
- \frac 1{m^2\, \mu^2} J^\alpha .
\end{eqnarray}
to write the second line in (\ref{LL100}).

Finally let us examine the contents of the matter equation in order to
compare with self-dual case.  The field equation for the scalar
field reads,

\begin{eqnarray}
\label{LL111}
& &\left(\partial_\mu\partial^\mu + M^2\right)\phi = \frac{e^2\, m^2}
{2\, \mu^4}\, \phi F_{\mu\nu}^2 + \frac{e^4}{\mu^4} \, \phi J_\mu^2\nonumber\\
& &-\frac{2\, i\, e^2}{\mu^2} J_\mu\partial^\mu\phi+
\frac{2\, e^3\,m^2}{\mu^4}\,\phi F_\mu\,J^\mu
-\frac{2\, i\, e\, m^2}{\mu^2} F_\mu\partial^\mu\phi\nonumber\\
& & = m^2\, e^4\, \phi\, \left( M_{\alpha\beta}  J^\beta\right)^2
-2\, i\, m\,e^2\,\partial^\alpha \phi\, M_{\alpha\beta} J^\beta
\end{eqnarray}
where we have used the vector field solution (\ref{LL100}) to get rid of
$F_\mu$ in the last line of the expression above and the property
(\ref{LL110}) repeatedly.
A simple inspection shows that this field equation has the same physical
contents as the matter equation of self-dual sector,
Eq.(\ref{LL20}).  This complete the proof that the bosonic matter
sector is unaltered by the duality mapping.

\section{Conclusions}

In this paper we studied the dual equivalence between the self-dual
model \cite{TPvN} and the Maxwell-Chern-Simons theory \cite{DJT} coupled
to dynamical matter, both fermionic and bosonic, using the iterative gauge
embedding procedure \cite{IW}. In the former case we reproduced the results
of \cite{GMdS} where the SD theory is mapped into the MCS theory and showed
that, (i) the dual mapping exchanges the minimal coupling between the fields
for a non-minimal magnetic interaction and (ii) introduces a current-current
Thirring-like interaction to preserve the dynamics of the fermionic matter
sector. The results of the latter case are new and, in fact, quite surprising.
We found that the dual transformation does not map the self-dual field into
a ``pure" MCS field but, instead, to a gauge field of the MCS type immersed
in a field dependent medium.  The current-current term and the magnetic
interaction were also induced by the dual mapping since their presence is
necessary to keep the matter dynamics unaltered but their coefficients were
found to be field dependent as well. This sort of model has been intensively
investigated \cite{P} and is known to give rise to vortex like solutions. 
\vspace{.5cm}

\noindent ACKNOWLEDGMENTS: This work is partially supported by CNPq, CAPES, FAPERJ and
FUJB, Brazilian Research Agencies.  The authors would like to thank D. Bazeia for many
suggestions.  CW thanks the Physics Department of UFPB the kind hospitality
during the course of this investigation.


\begin{thebibliography}{99}
\bibitem{CST} S. Chern and J. Simons, Ann. Math. 99 (1974) 49.
\bibitem{DJT} S. Schoenfeld, Nuc. Phys. B 185 (1981) 157;R. Jackiw and
S. Templeton, Phys. Rev. D23 (1981) 2291; S. Deser, R. Jackiw and
S. Templeton, Ann. Phys. 140 (1982) 372; Phys. Rev. Lett. 48 (1982) 372. 
\bibitem{boson} E.C. Marino, Phys. Lett. B263 (1991) 63; C.P. Burgess,
C.A. Lutken and F. Quevedo, Phys. Lett. B336 (1994) 18; C.P. Burgess and
F. Quevedo, Nucl. Phys. B421 (1994) 373; E. Fradkin and F. A. Schaposnik,
Phys. Lett. B338 (1994) 253;  R. Banerjee, Phys. Lett. B358 (1995) 297.
\bibitem{DJ} S. Deser and R. Jackiw, Phys. Lett. B 139 (1984) 2366.
\bibitem{TPvN} P. K. Townsend, K. Pilch and P. van Nieuwenhuizen, Phys.
Lett. B 136 (1984) 38.
\bibitem{GMdS} M. Gomes, L.C. Malacarne, A.J. da Silva, Phys. Lett.
B439 (1998) 137.
\bibitem{BFMS}N. Brali\'c, E. Fradkin, V. Manias and F. A. Schaposnik,
Nuc. Phys. B 446 (1995) 144;
\bibitem{IW} A. Ilha and C. Wotzasek, ``Duality Equivalence Between
Self-Dual And Topologically Massive Non-Abelian Models," hep-th/0012046.
\bibitem{footnote}{The Euler vectors $K_\mu$, are defined by the independent variations of
the action, whose kernel gives the equations of motion.}
\bibitem{vithee1}A.S. Vytheeswaran, ``Hidden symmetries in second class
constrained systems: are new fields necessary?," Workshop on Current
Topics in Quantum Field Theory, Calcutta, India, hep-th/0007230.
\bibitem{P}J. Lee and S. Nam, Phys. Lett. B {\bf261}, 437 (1991); D. Bazeia,
Phys. Rev. D {\bf46}, 1879 (1992); M. Torres, Phys. Rev. D 46 (1992) 2295.
\end{thebibliography}
\end{document}